\documentclass[useAMS,usenatbib]{mn2e}
\usepackage{epsfig}
\usepackage{amsmath}
\usepackage{rotating}

\newcommand{\be}{\begin{equation}}
\newcommand{\beq}{\begin{equation}}
\newcommand{\ba}{\begin{eqnarray}}
\newcommand{\ee}{\end{equation}}
\newcommand{\eeq}{\end{equation}}
\newcommand{\ea}{\end{eqnarray}}

\newcommand{\hs}{\hspace{1mm}}

\newcommand{\apj}{ApJ}

\newcommand{\apjl}{ApJL}
\newcommand{\mnras}{MNRAS}
\newcommand{\aj}{AJ}

\newcommand{\nat}{{\it Nature}}

\newcommand{\physrep}{Physics Reports}

\def\lsim{~\rlap{$<$}{\lower 1.0ex\hbox{$\sim$}}}

\def\gsim{~\rlap{$>$}{\lower 1.0ex\hbox{$\sim$}}}

\title[The 3-cm Fine-Structure Line from The EoR.]{On the Detectability of the Hydrogen 3-cm Fine-Structure Line from the Epoch of Reionization.}

\author[Dijkstra et al.]{M. Dijkstra\thanks{E-mail:mdijkstr@cfa.harvard.edu}, A. Lidz, J.R. Pritchard\thanks{Hubble Fellow}, L.J. Greenhill, D.A. Mitchell., S.M. Ord,\newauthor and R.B. Wayth\\
Astronomy Department, Harvard University, 60 Garden Street, Cambridge, MA 02138, USA}
\begin{document}

\date{\today}
\pagerange{\pageref{firstpage}--\pageref{lastpage}} \pubyear{2006}

\maketitle

\label{firstpage}
\begin{abstract}
A soft ultraviolet radiation field, $10.2$ eV$<h\nu<13.6$ eV, that
permeates neutral intergalactic gas during the Epoch of Reionization
(EoR) excites the $2p$ (directly) and $2s$ (indirectly) states of
atomic hydrogen. Because the $2s$ state is metastable, the lifetime of
atoms in this level is relatively long,  which may cause the $2s$
state to be overpopulated relative to the $2p$ state.  It has recently
been proposed that for this reason, neutral intergalactic atomic
hydrogen gas may be detected in absorption in its 3-cm fine-structure
line ($2s_{1/2}\rightarrow 2p_{3/2}$) against the Cosmic Microwave
Background out to very high redshifts.  In particular, the optical
depth in the fine-structure line through neutral intergalactic gas
surrounding bright quasars during the EoR may reach $\tau_{FS}\sim
10^{-5}$. The resulting surface brightness temperature of tens of
$\mu$ K (in absorption) may be detectable with existing radio
telescopes. Motivated by this exciting proposal, we perform a detailed
analysis of the transfer of Ly$\beta,\gamma,\delta,...$ radiation, and
re-analyze the detectability of the fine-structure line in neutral
intergalactic gas surrounding high-redshift quasars. We find that
proper radiative transfer modeling causes the fine-structure
absorption signature to be reduced tremendously to $\tau_{FS}\lsim
10^{-10}$. We therefore conclude that neutral intergalactic gas during
the EoR cannot reveal its presence in the 3-cm fine-structure line to
existing radio telescopes.
\end{abstract}

 
\section{Introduction}
\label{sec:intro}

It has recently been proposed that it may be possible to observe
neutral atomic hydrogen gas during the Epoch of Reionization (EoR), in
absorption against the Cosmic Microwave Background (CMB) in its 3-cm
fine-structure line \citep[$2s_{1/2}\rightarrow 2p_{3/2}$,][hereafter
S07]{Sethi07}.

The possible detectability of the 3-cm line derives from the fact that
the 2s state of atomic hydrogen is metastable \citep{BT40}. That is,
the average lifetime of an undisturbed atom in the 2s-state is
$t_{2s}=1/A_{2s1s}\sim 0.1$ s, after which it decays into the 1s state
by emitting 2 photons \citep{Spitzer51}. For comparison, the mean
life-time of atoms in the 2p state is $t_{2p}=1/A_{2p1s}\sim 10^{-9}$
s. Hence, when atoms are excited into their 2s and 2p states at
comparable rates, the 2s level is expected to be highly overpopulated
relative to the 2p level. This may result in a non-negligible optical
depth in the $2s_{1/2}-2p_{3/2}$ fine-structure line \citep[see
e.g.][]{Wild52}.

The detectability of this line has been discussed in the context of
HII regions in the local Universe \citep[e.g.][for a more detailed
discussion]{Dennison05}.  In local HII regions, the 2s level is
populated mainly following recombination, while both recombination and
photoexcitation by Ly$\alpha$ photons populate the 2p state. The
Ly$\alpha$ excitation rate of the 2p-state is limited by dust, and the
2s-state is typically heavily overpopulated in HII regions which
result in 3-cm optical depths of order $\tau_{FS} \sim 10^{-3}$
(Ershov 1987; Dennison et al, 2005, where the absorption is
measured relative to the free-free continuum that is produced in the
HII region). S07 have argued that during the EoR, bright quasars that
are luminous in the restframe soft-UV (10.2 eV  $<E_{\gamma}<$ 13.6
eV) may indirectly photoexcite the 2s state of neutral hydrogen gas in
the low density intergalactic medium to levels such that
$\tau_{FS}\sim 10^{-5}$ for CMB photons passing through these
regions\footnote{It may be surprising that the sparsely populated
$n=2$ levels of atomic hydrogen can produce a detectable opacity. One
can estimate $\tau_{FS}$ by comparing it to the Gunn-Peterson optical
depth $\tau_{GP}$:
$\frac{\tau_{FS}}{\tau_{GP}}=\frac{n_{2s}}{n_{1s}}\frac{A_{FS}\lambda_{FS}^3}{A_{2p1s}\lambda_{2p1s}^3}$
$\sim 20\frac{n_{2s}}{n_{1s}}$. Since $\tau_{\rm GP}\sim$ a few times
$10^5$ for a fully neutral IGM at $z=6$, $\tau_{FS}=10^{-5}$ only
requires that $\frac{n_{2s}}{n_{1s}}\sim 10^{-12}$. It will be shown
in \S~\ref{sec:sim} that these levels can be reached in regions around
high-redshift quasars.}. Absorption of these CMB photons would result
in an absorption feature with a brightness temperature of a few tens
of $\mu$K at $1.4$ Ghz, which may be detectable with existing dish
antennas such as the {\it Green Bank Telescope } (GBT), {\it Parkes}
and the most compact configuration of the {\it Australia Telescope
Compact Array} \citep[ATCA, e.g.][]{Ca08}. This result is important,
because it implies that neutral intergalactic atomic hydrogen gas that
exists during  EoR may reveal itself in a line that is different (and
possibly easier to detect) than the well studied 21-cm hyperfine
transition \citep[e.g.][]{SR90,LZ04,Furlanettorev}.
\begin{figure}
\vbox{\centerline{\epsfig{file=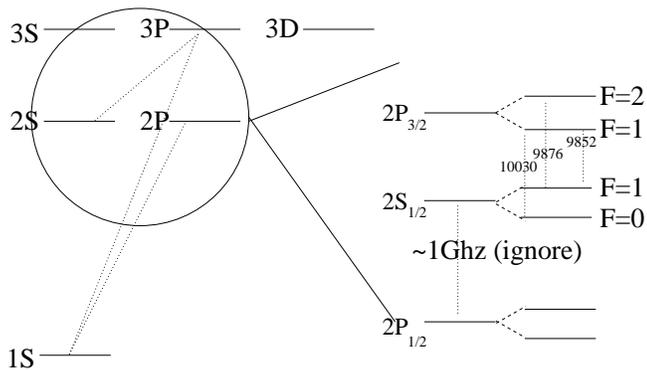,angle=0,width=8.5truecm}}}
\caption[]{Schematic diagram of the energy levels of a hydrogen
atom. On the left side, we show the first 3 energy levels, and their
splitting into levels with different angular momentum. The notation
for each level is $nL$, where $n$ is the principle quantum number, and
$L$ denotes the orbital angular momentum number ($L=S$ corresponds to
$l=0$, $L=P$ corresponds to $l=1$, $L=D$ corresponds to $l=2$). The
{\it dotted lines} show the allowed transitions through the emission
or absorption of a single photon. In a 3-level atom, the selection
rules only allow the 2s state to be populated by absorbing a Ly$\beta$
photon, and by subsequently emitting an H$\alpha$ photon. On the
right, the fine+hyperfine-structure splitting of the  $2s$ and $2p$
levels is shown in more detail. The Lamb-Rutherford  shift causes the
$2s_{1/2}$ level to lie above the $2p_{1/2}$-level by $\Delta
E/h_p\sim1.1$ Ghz \citep[e.g.][]{Wild52}. The 3 allowed transitions
between the $2s_{1/2}$ and  $2p_{3/2}$ levels are indicated with {\it
dashed lines}, and the frequency (in MHz) of each transition is
indicated \citep[also see][]{Dennison05}. The relative strength of
each transition 9852:9876:10030 is 1:5:2 \citep[][]{Wild52}.}
\label{fig:diagram}
\end{figure} 
Motivated by this exciting result, and by observational efforts to
detect this transition, we re-analyze the detectability of the
fine-structure line during the EoR. We perform more detailed modeling
of the transfer of Ly$\beta$ radiation than originally discussed by
S07. We will show that (unfortunately) proper modeling of Ly$\beta$
(and Ly$\gamma,\delta,...$) radiative transfer radically lowers the
value of $\tau_{FS}$ to levels that cannot be probed with existing
radio telescopes.
 
 The outline of this paper is as follows. We review the dominant
 excitation mechanisms of the 2s and 2p levels in \S\ref{sec:mech}. In
 \S\ref{sec:ex}, we compute the effective excitation rate of the
 2s-state during the EoR, by properly accounting for radiative
 transfer, and discuss the detectability of the 3-cm fine-structure
 line in \S\ref{sec:detection}. The validity of our model assumptions,
 and the implications of this work are discussed in
 \S\ref{sec:discuss}. Finally, we present the conclusions in
 \S\ref{sec:conc}. The parameters for the background cosmology used
 throughout are
 $(\Omega_m,\Omega_{\Lambda},\Omega_b,h)=(0.27,0.73,0.042,0.70)$
 \citep{Komatsu08}.

\section{The Excitation Mechanism of the 2s Level}
\label{sec:mech}

Figure~\ref{fig:diagram} depicts the first three principle energy
levels of the hydrogen atom. On the left, fine and hyperfine-structure
splitting of the lines are ignored. The notation that is used is $nL$,
where $n$ is the principle quantum number, and $L$ denotes the orbital
angular momentum number ($L=s$ corresponds to $l=0$, $L=p$ corresponds
to $l=1$, $L=d$ corresponds to $l=2$). In most astrophysical
conditions, the vast majority of hydrogen atoms are in the $1s$-state.
Atoms can be excited out of the ground state through the following
processes: ({\it i}) collisional excitation, ({\it ii}) recombination
following photoionization or collisional ionization, and ({\it iii})
photoexcitation. Following S07, processes ({\it i-ii}) are ignored
(see \S~\ref{sec:sim} - under roman numeral I -for a more detailed
motivation).

The quantum mechanical selection rules state that the only atomic
transitions that are allowed by emitting or absorbing a photon are
those in which $l$ changes by $\Delta l=\pm 1$. Hence, absorption only
results in transitions of the form $1s\rightarrow np$, while
subsequent emission only results in $np\rightarrow ms$. The allowed
transitions among the first 3 levels of the H-atom are indicated in
Figure~\ref{fig:diagram} as  {\it dotted lines}. In a 3-level atom
(the impact of using a multilevel atom is discussed in
\S~\ref{sec:sim} under roman numeral II) the selection rules only
allow the 2s state to be populated by absorbing a Ly$\beta$ photon,
and by subsequently emitting an H$\alpha$ photon. Hence, the level
population of atoms in the 2s state is determined by the {\it
effective Ly$\beta$ scattering rate}, denoted by $P_{\beta,{\rm
eff}}$, which is defined as the rate at which Ly$\beta$ photons
absorbed, and subsequently converted into H$\alpha$ + 2 continuum
photons. This rate is lower than to the "ordinary" Ly$\beta$
scattering rate, $P_{\beta}$, which is the rate at which Ly$\beta$
photons are absorbed.

\section{The 2s and 2p Level Populations}
\label{sec:ex}

As mentioned above, the effective Ly$\beta$ scattering rate,
$P_{\beta,{\rm eff}}$, is crucial in calculating the number of atoms
in the 2s-state. The goal of this section is to evaluate
$P_{\beta,{\rm eff}}(r)$. In \S~\ref{sec:mc} we describe the
Monte-Carlo technique that was used to compute the general and
effective Ly$\beta$ scattering rates, and in \S~\ref{sec:result} we
present our results.

\subsection{Outline of Monte-Carlo Calculation}
\label{sec:mc}

We compute the general and effective Ly$\beta$ scattering rate by
performing Monte-Carlo (MC) calculation of the Ly$\beta$ radiative
transfer in the neutral gas surrounding a spherical HII region of
radius $R_{\rm HII}=5$ Mpc, that is based on the physical size of the
HII region that possibly exist around $z=6$ quasars \citep[][but see
Lidz et al 2007, Bolton \& Haehnelt 2007]{MH04,WL04}. This MC-approach
to radiative transfer, in which the trajectories of individual photons
are followed, is common for Ly$\alpha$ radiation
\citep[e.g.][]{LR99,Ahn00,Zheng02}. We use the Ly$\alpha$-MC code
originally described in \citet{MC}, which is easily adapted for
Ly$\beta$ radiative transfer. The calculation is performed as follows:

{\bf 1.} A photon is emitted with a frequency that is drawn randomly
from the range $\nu\in[0.95\nu_{\beta},1.05\nu_{\beta}]$. The transfer
calculation therefore focuses on photons in a frequency range $\Delta
\nu_{\beta}=0.1\nu_{\beta}$ centered on $\nu_{\beta}$, similar to the
optically thin calculation that was originally performed by S07 (and
which is discussed in Appendix~\ref{sec:lya1}).

{\bf 2.} The photon propagates an optical depth $\tau=-\log R$, where
$R$ is a random number between 0 and 1. The optical depth translates
to a physical distance via $\tau=\int_0^Rds\hs n_{\rm
HI}(s)\sigma_{\beta}(\nu[1-Hs/c])$.

{\bf 3.} Each time a Ly$\beta$ photon is absorbed, there is a
$p_{\beta,{\rm scat}}=0.88$ probability for the atom to decay directly
back to the ground state by emitting a Ly$\beta$ photon
\citep{Hirata06,PF06}. This is simulated in the MC-calculation by
generating a random number $R\in[0,1]$. The photon scatters as a
Ly$\beta$ photon if $R\leq p_{\beta,{\rm scat}}$, in which case a new
direction \footnote{The scattering process is assumed to be partially
coherent, and described by a Rayleigh phase function, (see Dijkstra \&
Loeb 2008 for a more detailed discussion of the scattering phase
function).} is chosen and we return to step {\bf 2}. On the other
hand, a transition of the form $3p\rightarrow 2s$ occurs when
$R>p_{\beta,{\rm scat}}$. After this process, the Ly$\beta$ photon is
destroyed, and we return to step {\bf 1.}.

We compute total Ly$\beta$ scattering rate by keeping track of the
number of scattering events that occur in shell $j$, $N_j$ (in the
MC-simulation, the gas is sampled by a discrete number of gas shells,
see Dijkstra et al 2006). The total Ly$\beta$ scattering rate transfer
in shell $j$ follows from $P_{\beta,j}=\dot{N}_{\beta}\times
N_j/(N_{\rm H,j}\times N_{\rm mc})$, where $\dot{N}_{\beta}$ is the
total rate at which the quasar is emitting photons in the frequency
range $\nu\in[0.95\nu_{\beta},1.05\nu_{\beta}]$ (we have assumed
$\dot{N}_{\beta}=10^{58}$ s$^{-1}$ to compare to S07 as closely as
possible), $N_{\rm H,j}$ is the total number of hydrogen atoms in
shell $j$, and where $N_{\rm mc}$ is the total number of photons that
is used in the MC-simulation. The effective Ly$\beta$ scattering rate
is obtained similarly from $P_{\beta,j,{\rm
eff}}=\dot{N}_{\beta}\times N_{j,{\rm 3p2s}}/(N_{\rm H,j}\times N_{\rm
mc})$, where $N_{j,{\rm 3p2s}}$ is the total number of scattering
events of the sequence $1s\overset{{\rm
Ly}\beta}{\rightarrow}3p\overset{{\rm H}\alpha}{\rightarrow}2s$ in
shell $j$.

\subsection{The General and Effective Ly$\beta$ Scattering Rates derived from the MC-Simulation}
\label{sec:result}

\begin{figure}
\vbox{\centerline{\epsfig{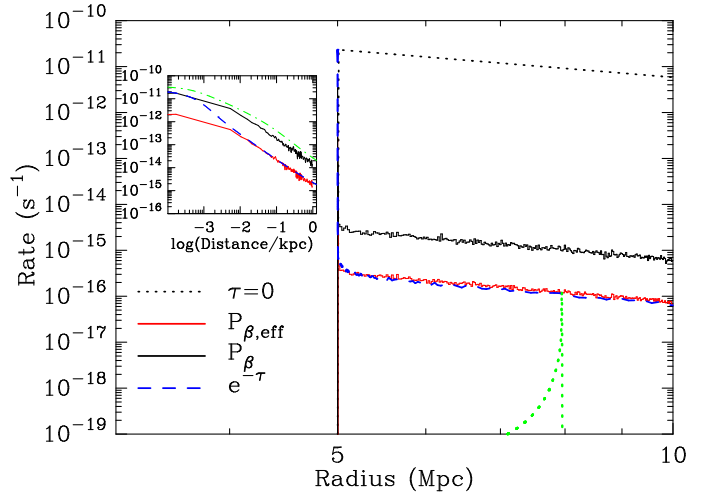}}}
\caption[]{The effective (general) Ly$\beta$ scattering rate that was
derived from the MC-simulation is plotted as the {\it red} ({\it
black}) {\it histogram}. The inset zooms in on the edge of the HII
region. Also shown as the {\it blue dashed line} is the scattering
rate that one obtains by suppressing the Ly$\beta$ mean intensity by a
factor of e$^{-\tau}$ (\S~\ref{sec:noscat}). The {\it green dot-dashed
line} (inset only) includes the contribution from higher Lyman series
lines (see \S\ref{sec:sim}). The {\it dotted line} shows the
scattering rate in the optically-thin limit. This Figure shows that:
({\it i}) the effective Ly$\beta$ scattering rate derived from the
MC-simulation is reproduced well by simply suppressing the mean
Ly$\beta$ intensity by a factor of e$^{-\tau}$ (see
\S~\ref{sec:noscat}); and ({\it ii}) the effective scattering rate
decreases rapidly with radius, (see inset) and levels off at a level
that is $\sim 5$ orders of magnitude smaller than the rate obtained
from the optically thin calculation. Lastly, the {\it green dotted
line} in the main plot shows the scattering rate of photons that would
have redshifted into resonance at radius=8 Mpc (see text)}
\label{fig:lybscatrate}
\end{figure}

The general and effective Ly$\beta$ scattering rates we obtain from
the Monte-Carlo calculation are shown in Figure~\ref{fig:lybscatrate}
as the {\it black} and {\it red solid histograms}, respectively. The
{\it blue dotted line} shows the scattering rate that is obtained when
the gas is assumed to be optically thin (which corresponds to the
original calculation of S07, see Appendix \ref{sec:lya1} for more
details). Furthermore, the {\it blue dashed line} shows the scattering
rate that one obtains if one ignores that Ly$\beta$ photons undergo
multiple scattering events, and simply suppresses the incoming
Ly$\beta$ flux by a factor of e$^{-\tau}$ (this calculation is
described in Appendix \ref{sec:noscat}).

Figure~\ref{fig:lybscatrate} shows that the effective Ly$\beta$
scattering rate is much lower than the rate that is obtained in the
optically thin calculation. At small distances into the HI region
($d>10^{-2}$ kpc, which is shown in the inset), the effective
scattering rate is $P_{\beta,{\rm eff}}<10^{-13}$ s$^{-1}$. The
effective scattering rate decreases rapidly with radius, and levels
off at a level that is $\sim 5$ orders of magnitude smaller than the
rate one obtains from the optically thin calculation. This rapid
decrease owes to the large line-center optical depth of the neutral
IGM to Ly$\beta$ photons, which causes resonant scattering only to
occur near the edge of the HII region. This point is illustrated by
the {\it green dotted line} in the main plot of
Figure~\ref{fig:lybscatrate}, which shows the scattering rate of
photons with frequencies such that they would have redshifted into
resonance at $r=8$ Mpc (i.e. 3 Mpc into the neutral IGM). When these
photons reach $r=7$ Mpc, they find themselves $\sim 750$ km s$^{-1}$
blueward of the resonance, and their absorption cross-section is
suppressed by a factor of  $\sim 10^{-8}$ relative to the resonance
absorption cross-section. The scattering rate of these photons is
therefore also suppressed by a factor of $10^{-8}$ (relative to the
optically thin rates). The scattering rate increases rapidly with
radius as the photons redshift closer to resonance, and the maximum
scattering rate occurs at $r=7.95$ Mpc, where the photons have
redshifted to within $\sim 3$ Doppler widths from resonance. The
optical depth of the IGM increases rapidly as these photons redshift
closer to resonance, which results in a sharp cut-off in the
scattering rate at $r\gsim 7.95$ Mpc.   In other words, photons that
were emitted somewhat blueward of the Ly$\beta$ resonance frequency
-which need to travel far into neutral gas from the HII region edge
before redshifting into the line center- have a high probability of
being absorbed in the damping wing of the line, and do not generally
make it to line center. The absorption cross section in the damping
wing is low compared to that at line center, and so these photons
yield a correspondingly small scattering rate. Therefore the Ly$\beta$
scattering rates are strongly suppressed at $d\gsim10^{-2}$
kpc\footnote{Note that in our model the total scattering rate, which
is given by $\mathcal{P}_{\rm tot}\equiv n_H\int_{R_{\rm
HII}}^{\infty}dr\hs 4\pi r^2\hs P_{\beta,{\rm
eff}}(r)=\dot{N}_{\beta}$, i.e. photons are conserved. On the other
hand, $\mathcal{P}_{\rm tot}\gg\dot{N}_{\beta}$ in the optically thin
limit (\S\ref{sec:lya1}). Each emitted Ly$\beta$ photon can only
undergo one effective scattering event, and the total effective
scattering rate can never exceed the total rate at which Ly$\beta$
photons are emitted. The optically thin calculation is therefore
logically inconsistent.}. The sharp cut-off of the effective Ly$\beta$
scattering rate is also discussed in Appendix \ref{sec:noscat}.

Furthermore, Figure~\ref{fig:lybscatrate} shows that the effective
scattering rate that is extracted from the MC-simulation ({\it red
solid histogram}) is well reproduced by the model in which the mean
Ly$\beta$ intensity is suppressed by a factor of e$^{-\tau}$ (the {\it
blue dashed line}). The reason for this is that each Ly$\beta$ photon
can only excite 1 atom into the 2s state: each time a transition of
the from
$1s\overset{Ly\beta}{\rightarrow}3p\overset{H\alpha}{\rightarrow}2s$
occurs, the Ly$\beta$ photon is lost permanently. This almost
corresponds to a calculation in which one ignores scattering and
simply suppresses the intensity with a factor e$^{-\tau}$. The only
difference is that Ly$\beta$ photons typically scatter $\langle N_{\rm
scat} \rangle=1/(1-p_{\beta,{\rm scat}})\sim 8.3$ times before being
converted into H$\alpha$ + 2 continuum photons. The random walk
associated with the scattering process transports the Ly$\beta$
photons on average $\sim \sqrt{\langle N_{\rm scat} \rangle}\sim 2.9$
mean free paths through the neutral IGM, before indirectly exciting an
atom into its 2s-state. In practise, this introduces a difference with
the scattering rate obtained from the e$^{-\tau}$-calculation only at
the smallest separations from the edge of the HII region ($r<10^{-2}$
kpc, see inset). At these smallest radii, the gas is optically thin to
Ly$\beta$ photons, and here the {\it total} Ly$\beta$ scattering rate
({\it black solid histogram}) as extracted from the MC-simulation
agrees best with the analytic solution given by Eq~(\ref{eq:pscat2})
with $\tau=0$. Naturally, here the {{\it effective}} scattering rate
is suppressed by a factor of $(1-p_{\beta,{\rm scat}})=0.12$.

\section{The Signal}
\label{sec:detection}

The change in brightness temperature for CMB radiation passing through
the neutral gas with an enhanced population of atoms in the 2s-state
is given by

\begin{equation} 
\Delta T_b=\frac{\tau_{FS}}{1+z}(T_{\rm ex}-T_{\rm CMB}),
\label{eq:tb}
\end{equation} where $T_{\rm CMB}=2.73(1+z)$ is the CMB temperature at redshift $z$, and $T_{\rm ex}$ is the excitation temperature of the 2s-2p transition, which is defined as $T_{\rm ex}\equiv\lambda_{FS}^2S/2k_B$, where $S=\frac{2h\nu^3_{FS}}{c^2}\frac{n_u}{\frac{g_u}{g_l}n_l-n_u}$ is the standard source function \citep[see][Eq~1.79]{RL79}, $n_l$ ($n_u$) denotes the number density of atoms in the 2s level (2p level), and $g_l$ ($g_u$) denotes the statistical weight of this level. 
Throughout we ignore that hyperfine splitting of the 2s and 2p levels
actually results in 3 possible 2s-2p transitions (see
Fig~\ref{fig:diagram}). We discuss the impact of hyperfine splitting
on our results in \S~\ref{sec:sim} (under roman numeral
V). Furthermore, the optical depth to 2s-2p absorption, $\tau_{FS}$ is
given by \citep[][Eq 1.78]{RL79}

\begin{equation} 
\tau_{FS}(\nu)=\int ds\hs \alpha(\nu(s))=\frac{h\nu_{FS}B_{ul}}{4\pi
\Delta \nu_{FS}}\int
ds\hs\Big{(}\frac{g_u}{g_l}n_l-n_u\Big{)}\phi(\nu(s)),
\label{eq:tau2s}
\end{equation} where $\Delta \nu_{FS}=v_{\rm th}\nu_{FS}/c$, $B_{ul}$ is the Einstein B-coefficient for the fine-structure transition, and $\phi(\nu)$ is the line profile function \citep[see][]{RL79}. S07 move the factor $\frac{g_u}{g_l}n_l-n_u$ outside of the integral and arrive at their Eq~(7). However this is not allowed, because we found $\Gamma_{\beta}(r)$, and therefore $n_{l}(=n_{2s})$, to be a rapidly changing function of radius (see Fig~\ref{fig:lybscatrate}). This makes the evaluation of $\tau_{FS}(\nu)$ somewhat trickier. Fortunately, we may simplify Eq~\ref{eq:tb} and Eq~\ref{eq:tau2s} by assuming that $n_l> n_u$ (which is verified in \S~\ref{sec:sim}, under roman numeral IV). Under this assumption, $T_{\rm ex}\ll T_{\rm CMB}$ and $\Delta T_b(\nu)=-\tau_{FS}(\nu)T_{\rm CMB}/(1+z)$, and the optical depth to 2s-2p absorption at frequency $\nu$ simplifies to
\begin{equation}
\tau_{FS}(\nu)=\int ds\hs n_{2s}(s)\sigma_{FS}(\nu[1-Hs/c]),
\end{equation} where  $\sigma_{FS}(\nu)=\frac{h\nu_{FS}B_{ul}}{4\pi \Delta \nu_{FS}}\frac{g_u}{g_l}\phi(\nu)$. For simplicity, we assume that the edge of the HII region is infinitely sharp (i.e. the HII region contains no neutral hydrogen, and the neutral fraction jumps from $0$ to $1$ instantly at the edge of the HII region) and extends out to $r=R_{\rm HII}=5$ Mpc (see \S~\ref{sec:sim}, roman numeral III). Under this assumption, the optical depth along a path that is pointing radially outward from the quasar is given by 
\begin{equation}
\tau_{FS}(\nu)=\frac{n_{\rm H}}{A_{2s1s}}\int_{R_{\rm HII}}^{\infty}
dr\hs P_{\beta}(r)\sigma_{FS}(\nu[1-Hr/c]),
\label{eq:tau2sf}
\end{equation} where we used that $n_{2s}=\frac{n_{1s}P_{\beta}}{A_{2s1s}}=\frac{n_{H}P_{\beta}}{A_{2s1s}}$. The last approximation is allowed because the vast majority of all hydrogen atoms are in the 1s state. 

The life-time of the 2p state is set by the Einstein coefficient of
the $2p\rightarrow1s$ transition, $A_{2p1s}=6.25\times 10^8$ s$^{-1}$,
and the line profile of the 2s-2p transition has a Voigt parameter $a=
A_{2p1s}/(4\pi\Delta \nu_{FS})=117(T_{\rm gas}/10^4\hs{\rm K})^{-1/2}$
\citep{Dennison05}. The line profile is therefore completely dominated
by the damping wing, and the cross-section displays a much weaker
frequency dependence than is common among the Lyman-series
transitions: e.g. even for $x_{FS}=100$, where
$x_{FS}(\nu)=(\nu-\nu_{FS})/\Delta \nu_{FS}$, the cross-section is
only reduced by a factor of $\phi(x_{FS})/\phi(0)=0.57$.

\begin{figure}
\vbox{\centerline{\epsfig{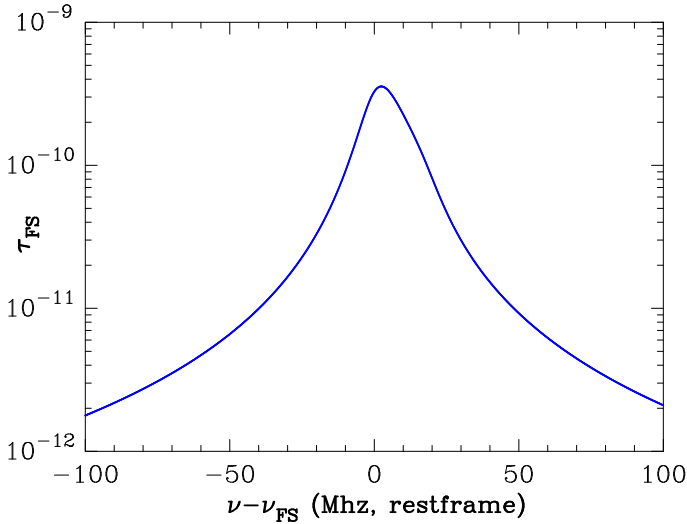}}}
\caption[]{This Figure shows the optical depth in the 2s-2p
fine-structure transition, $\tau_{FS}(\nu)$, through neutral gas
surrounding a quasar HII region at $z=6.5$. The maximum optical depth
is $\tau_{FS{\rm,max}}\sim 4 \times 10^{-10}$, which results in a
maximum brightness temperature of $\sim 10^{-10}$ K. This brightness
temperature if five orders of magnitude lower than previous estimates,
and likely precludes detecting hydrogen fine-structure absorption
during the EoR.}
\label{fig:tau}
\end{figure}

We compute $\tau_{FS}(\nu)$ by using $\Gamma_{\beta}(r)$ we found in
\S~\ref{sec:result} (and we account for the additional contribution of
higher Lyman series following \S\ref{sec:sim}) into
Eq~\ref{eq:tau2sf}. The result of this calculation is shown in
Figure~\ref{fig:tau},
 which shows that the maximum optical depth is $\tau_{FS,{\rm max}}=4\times 10^{-10}$. This is 5 orders of magnitude smaller than the value that was found by S07. The reason for this large difference is the assumption that the factor $\frac{g_u}{g_l}n_l-n_u$ could be taken outside the integral in Eq~\ref{eq:tau2s}. This is allowed only when the dependence of this factor on $r$ is weaker than that of $\sigma_{FS}(\nu[1-Hr/c])$. However, because the 2s-2p absorption cross-section is completely dominated by the damping wing, it has a tremendous FWHM that is $\nu_{\rm FWHM}/\nu_{FS}\sim 0.01$. At $z=6.5$, Hubble expansion over 4 Mpc introduces a comparable frequency shift of $\Delta \nu/\nu\sim 0.01$, and S07 therefore implicitly assumed that their calculated level population in the 2s-state applied to a 'shell' that was at least $4$ Mpc thick. In sharp contrast, we found that their calculation applied only to a shell of gas that was $\sim 10^{-2}$ kpc thick (see the inset of Fig~\ref{fig:lybscatrate}). In other words, the total column of neutral H atoms in the 2s state is lower by $\sim$5 orders of magnitude.

 It is worth emphasizing that our calculation of $\tau_{FS,{\rm max}}$
 assumed that the quasar emitted $\dot{N}_{\beta}=10^{58}$ photons
 s$^{-1}$ in the frequency range $\nu \in[0.95,1.05]\nu_{\beta}$. This
 is very likely a factor of $\sim 10$ too high (see Appendix
 \ref{sec:lya1}). Therefore, in reality the value of  $\tau_{FS,{\rm
 max}}$ is likely even lower by an additional factor of $\sim 10$.

\section{Discussion}
\label{sec:discuss}

\subsection{Validity of Model Assumptions}
\label{sec:sim}

\hspace{6mm}(I): In \S\ref{sec:mech} we mentioned that the collisional
excitation rate of -and the recombination rate into- the 2s state,
were negligibly small compared to the effective Ly$\beta$ scattering
rate. For example, the collisional excitation rate is given by
$C_{2s}=2.24\times 10^{-6}n_eT_{\rm gas}^{-1/2}\exp(-10.2\hs{\rm
eV}/kT_{\rm gas})$ \citep[][p54-55]{Osterbrock89}. If we assume that
the gas is half ionized ($n_e=0.5n_{\rm H}$), then we find that
$C_{2s}< 10^{-14}$ s$^{-1}$ for $T_{\rm gas}<3\times 10^4$ K. This
rate is orders of magnitude less than the effective Ly$\beta$
scattering rates in the regions of interest. Similarly, we find that
the recombination rate is $R_{2s}<\alpha_{\rm rec,B}n_e=10^{-17}$
s$^{-1}$, which is even smaller.

(II): So far we have ignored the fact that the 2s state of atomic
hydrogen can also be accessed following the absorption of higher order
Lyman series photons, e.g. via a scattering event of the form
$1s\overset{{\rm
Ly}\delta}{\rightarrow}5p\overset{H\gamma}{\rightarrow}2s$. In
principle, it is straightforward to expand the MC-approach to include
scattering of these higher-order Lyman series photons, but this is not
necessary. We found in \S~\ref{sec:result} that the effective
Ly$\beta$ scattering rate was calculated reasonably well if one simply
suppresses the incoming Ly$\beta$ flux with a factor of
e$^{-\tau}$. For simplicity, we also use this approach to calculate
the effective scattering rates of the higher order Lyman series
photons, and cast the total effective scattering rate in a form
similar to that of Eq~(\ref{eq:pscat2}):
\begin{equation}
P_{\rm eff,tot}=\frac{L_{\nu}\pi e^2}{4\pi r^2\sqrt{\pi}m_e
c}\sum_{n=\beta,\gamma,...}\frac{f_{n}}{h\nu_{n}}\int
dx_n\phi_n(x_n){\rm e}^{-\tau(x_n,r)},
\end{equation} where $f_n$ is the oscillator strength, $h\nu_{n}$ is the energy, and $\phi_n(x_n)$ is the Voigt function\footnote{The oscillator strengths can be found Chapter 10.5 in Rybicki \& Lightman (1979). The Einstein coefficients, and therefore the absorption cross-sections, can be derived from the oscillator strength using Eq~10.79 of Rybicki \& Lightman 1979.} for the Lyman-n transition (i.e. $1s\rightarrow np$), where $x_n=(\nu-\nu_n)/\Delta \nu_D$. The first term of the series corresponds to Eq~(\ref{eq:pscat2}). The result of summing the first 10 Lyman series transitions is shown as the {\it green dot-dashed line} in Figure~\ref{fig:lybscatrate}. Including these higher Lyman series photons only boosts the effective scattering rates by a factor of $\sim 1.5$ in the region that was optically thin to Ly$\beta$ photons, which is due to the fact that the oscillator strength of $np\rightarrow 1s$ transitions -and hence the absorption cross section- decreases quite rapidly with $n$ (e.g. for $10p\rightarrow 1s$, $f_{10}=0.0016$, which is $\sim 0.004f_{\alpha}$). Between $10^{-2}-10^{-1}$ kpc from the edge of the HII region, the contribution from higher Lyman series photons is the largest and boosts the overall scattering rate by a factor $\sim 15$. This larger boost is due the fact that at these distances the gas is not yet optically thick in these higher Lyman series lines. Note that we have accounted for this boost in our calculation of $\tau_{FS}$ (which is shown Figure~\ref{fig:tau}).

(III): We have so far assumed that the edge of the HII region is
infinitely sharp. In reality, the thickness of the HII region is
determined by the mean free path of the ionizing photons emitted by
the quasar. Especially the X-Rays emitted by the quasar can penetrate
deeply into the neutral IGM and produce a low level of
ionization. However, the presence of a partially neutral region does
not affect our results at all: although the distance from the edge of
the HII region at which the IGM becomes optically thick is increased
by a factor of $1/x_{\rm HI}$ (where $x_{\rm HI}$ is the neutral
fraction), $\tau_{FS}$ is reduced by the same factor.

(IV): The approximation $\Delta T_b(\nu)=-\tau_{FS}(\nu)T_{\rm
CMB}/(1+z)$ is valid only when $T_{\rm ex}\ll T_{\rm CMB}$, or when
$n_l > n_u$. The 2s and 2p populations are given by
$n_{2s}=n_l=n_H\Gamma_{\beta}/(A_{2s1s})$ and
$n_{2p}=n_u=n_H\Gamma_{\alpha}/(A_{2p1s})$, respectively. That is,
$n_l > n_u$ when
$\Gamma_{\beta}>(A_{2s1s}/A_{2p1s})\Gamma_{\alpha}\sim
10^{-8}\Gamma_{\alpha}$. Our assumption $n_l > n_u$ therefore breaks
down in the regions of interest only when $\Gamma_{\alpha}\gsim
10^{-4}$ s$^{-1}$, which is orders of magnitude larger than the actual
Ly$\alpha$ scattering rate.

(V): Our calculation has ignored the hyperfine splitting of the 2s and
2p levels that was illustrated in Figure~\ref{fig:diagram}. Hyperfine
splitting results in 3 allowed 2s-2p transitions at $\nu=9.852,9.876$
and $10.030$ Ghz, and the fine-structure absorption cross-section has
maxima at these frequencies. The most prominent of these 3 maxima
reaches $75\%$ of the value that one obtains when hyperfine-structure
splitting is ignored \citep[][]{Dennison05}. Including
hyperfine-structure therefore lowers the maximum observable brightness
temperature by a factor of $0.75$.

\subsection{General Detectability of the Fine-structure line during the EoR}

We have shown that the optical depth in the 2s-2p fine-structure line
 around quasar HII regions to be negligible because of the low
 effective Ly$\beta,\gamma,\delta,...$ scattering rates. Early during
 the EoR, the IGM was exposed to a background Lyman series photons
 generated by early generations of galaxies before it was
 significantly ionized \citep[e.g.][]{Furlanettorev}. In principle,
 one might imagine looking for fine-structure absorption in the IGM
 during this early phase of the EoR.  Unfortunately, the effective
 Ly$\beta$ scattering rate during this phase of the EoR will be low
 for the same reasons it is low around bright quasars, as we
 considered previously:  it is well established that Ly$\alpha$
 scattering rate plays an important role in determining the excitation
 temperature of the 21-cm hyperfine transition via the
 Wouthuysen-Field effect \citep[see
 e.g.][]{Madau97,Furlanettorev}. However, each soft-UV photon that
 redshifts toward (and through) the Ly$\alpha$ resonance frequency
 scatters on average $N_{\rm scat}\sim \tau_{\rm GP}\sim 10^5-10^6$
 times. In sharp contrast, each higher Lyman series photon only
 excites one atom into the $2s$-state. The effective Ly$\beta$ (and
 total) scattering rate is therefore lower than the Ly$\alpha$
 scattering rate by a factor of $\sim \tau_{\rm GP}$. Since the
 Ly$\alpha$ scattering rate in the IGM during the EoR is typically
 $P_{\alpha}\sim 10^{-12}-10^{-11}$ s$^{-1}$ \citep[e.g.][]{Madau97},
 the effective Ly$\beta$ scattering rate is of the order
 $P_{\beta,{\rm eff}}\sim 10^{-18}-10^{-16}$ s$^{-1}$, which barely
 competes with the recombination rate. From this we conclude that
 wherever the Wouthuysen-Field effect operates in the low density
 neutral IGM, the optical depth in the fine-structure line is expected
 to be $\tau_{FS}<10^{-10}$.

\section{Conclusions}
\label{sec:conc}
It has recently been proposed that neutral intergalactic atomic
hydrogen gas may be detected in absorption in its 3-cm fine-structure
line ($2s_{1/2}\rightarrow 2p_{3/2}$) against the Cosmic Microwave
Background (CMB) out to very high redshifts. In particular, bright
quasars that are luminous in the restframe soft-UV (10.2 eV
$<E_{\gamma}<$ 13.6 eV) may indirectly photoexcite the 2s state of
neutral hydrogen gas in the intergalactic medium (IGM) during the
Epoch of Reionization (EoR) to levels such that $\tau_{FS}\sim
10^{-5}$ for CMB photons passing through these regions
\citep{Sethi07}. The resulting brightness temperature of $\Delta
T_b\sim$ tens of $\mu$K could be detected with existing radio
telescopes \citep[e.g.][]{Ca08}.

Motivated by this proposal, and by observational efforts to detect
this transition, we have performed a detailed analysis of the transfer
of Ly$\beta,\gamma,\delta,...$ radiation, and have re-analyzed the
detectability of the fine-structure line in neutral intergalactic gas
surrounding bright quasars during the EoR. We have found that properly
considering radiative transfer effects reduces the signal
significantly compared to optically thin estimates, which radically
complicates the detectability of the 3-cm fine structure transition.

The main reason for this negative result is the large opacity of the
(partially) neutral IGM to Lyman series photons. In such a medium,
Ly$\alpha$ scattering proceeds completely different than
Ly$\beta,\gamma,\delta,...$ scattering: Ly$\alpha$ photons scatter
$\tau_{\rm GP}\sim10^5-10^6$ times.   However, higher Lyman series
photons have a finite probability for being converted into Balmer,
Paschen, .., etc photons at each scattering event, which subsequently
propagate to the observer unobstructed. These higher Lyman series
photons therefore scatter on average only $5-8$ times
(\S\ref{sec:result}) before being destroyed \citep{PF06}. Furthermore,
only one of these $5-8$ scattering events - defined as an {\it
effective scattering event}- indirectly excites atoms into their
2s-state (e.g. Ly$\beta$ scattering corresponds to a sequence
$1s\rightarrow 3p\rightarrow 1s$, and does not affect the level
population of the $2s$-state). For these reasons, we found that the
effective Ly$\beta$ scattering rate was calculated reasonably well if
one simply suppresses the incoming Ly$\beta$ flux by a factor of
e$^{-\tau}$, in which $\tau$ is the (frequency dependent) optical
depth of the IGM to Ly$\beta$ radiation (see
Fig~\ref{fig:lybscatrate}). The same applies to Ly$\gamma,\delta,...$
radiation.

Because of the reduced effective scattering rates, we found a
substantial lower column of neutral atoms in the 2s-state. This, in
combination with the fact that the profile of the fine-structure
absorption cross section is dominated by the damping wings
(\S~\ref{sec:detection}), results in $\tau_{FS}\sim
10^{-11}-10^{-10}$. This is 5-6 orders of magnitude smaller than what
was originally found by \citet{Sethi07}, and we conclude that the 3-cm
fine-structure absorption line from neutral intergalactic gas
surrounding high-redshift quasars is presently undetectable.

{\bf Acknowledgments} This research was supported by Harvard
University Funds. JRP is supported by NASA through Hubble Fellowship
grant HST-HF-01211.01-A awarded by the Space Telescope Science
Institute, which is operated by the Association of Universities for
Research in Astronomy, Inc., for NASA, under contract NAS 5-26555. SMO
is supported by NSF/AST0457585.

\appendix
\section{Analytic Calculations of Ly$\beta$ Scattering Rates}
\subsection{The Optically Thin Limit}
\label{sec:lya1}

 In the regime where the gas is optically thin to Ly$\beta$ radiation,
 the Ly$\beta$ scattering rate is given by

\begin{equation}
P_{\beta}=4\pi\int d\nu\hs\frac{J(\nu)}{h\nu}\sigma_{\beta}(\nu),
\label{eq:pscat}
\end{equation} where $J(\nu)$ is the mean intensity in the UV radiation field near the Ly$\beta$ resonance (in erg s$^{-1}$ cm$^{-2}$ Hz$^{-1}$ sr$^{-1}$), and $\sigma_{\beta}(\nu)$ is the Ly$\beta$ absorption cross-section. The mean intensity $J(\nu)$ relates to the luminosity of the central source (in erg s$^{-1}$ Hz$^{-1}$) as $J(\nu)=\frac{L_{\nu}}{16\pi^2r^2}$, where $r$ is the distance to the central source. Assuming that $J(\nu)$ is constant across the Ly$\beta$ resonance, which is extremely narrow in frequency, Eq~\ref{eq:pscat} simplifies to

\begin{equation}
P_{\beta,{\rm thin}}=\frac{L_{\nu}}{4\pi r^2\hs h\nu_{\beta}}\int
d\nu\hs\sigma_{\beta}(\nu),
\end{equation} where the photon energy $h\nu$, that was originally in the integrand, was taken outside of the integral since it barely varies over the limited range of frequencies where the integrand is not negligibly small. Using that $\int d\nu\hs\sigma_{\beta}(\nu)=f_{\beta}\frac{\pi e^2}{m_e c}$, where $f_{\beta}=0.079$ is the Ly$\beta$ oscillator strength, $c$ is the speed of light, $e$ and $m_e$ are the electron charge and mass, respectively \citep[see e.g.][their Eq 3.66]{RL79}, the total Ly$\beta$ scattering rate simplifies further to

\begin{equation}
P_{\beta,{\rm thin}}=\frac{L_{\nu}\hs f_{\beta}\hs \pi e^2}{4\pi
r^2\hs h\nu_{\beta}\hs m_e c}.
\label{eq:pscatb}
\end{equation} The only quantity that remains to be determined is $L_\nu$. Under the assumption that $L_{\nu}$ does not vary with frequency near the Ly$\beta$ resonance frequency, $L_\nu=L/\Delta \nu$, where $L$ is the total luminosity (in erg s$^{-1}$) of the central source between frequency $\nu_{\rm min}$ and $\nu_{\rm max}$, and $\Delta \nu \equiv (\nu_{\rm max}-\nu_{\rm min})$. S07 assume that\footnote{This assumption corresponds to a flux density that is $6\times 10^{32}$ erg s$^{-1}$ Hz$^{-1}$, which corresponds to an absolute AB magnitude of $M_{AB}=-30.45$. For comparison, the brightest $z=6$ quasars have $M_{1450}\sim -27.8$ \citep{Fan06}. Given that quasars have a relatively flat spectrum between $\lambda=\lambda_{\beta}$ and $\lambda=1450$ \AA\hs (rest-frame), we conclude that the quasar luminosity assumed by S07 is likely too high by about a factor of $\sim 10$.} $\dot{N}_{\beta}=10^{58}$ photon s$^{-1}$ and $\Delta \nu\sim 0.1\nu_{\beta}$, and obtain
\begin{equation}
P_{\beta,{\rm thin}}(r)=2.4\times 10^{-11}\hs{\rm
s}^{-1}\times\Big{(}\frac{5\hs{\rm Mpc}}{r}\Big{)}^2,
\end{equation} where we used that $L=\dot{N}_{\beta}\times h\nu_{\beta}=1.9\times 10^{47}$ erg s$^{-1}$, and $r=5$ Mpc (\S~\ref{sec:mc}). This scattering rate was plotted as the {\it dotted line} in Figure~\ref{fig:lybscatrate}.

\subsection{Including Radiative Transfer: Ignore Scattering}
\label{sec:noscat}

When the assumption of working in the optically thin regime is
dropped, the scattering rate (Eq~\ref{eq:pscat}) modifies to

\begin{equation}
P_{\beta}(r)=4\pi\int
d\nu\hs\frac{J(\nu)}{h\nu}\sigma_{\beta}(\nu[1-Hr/c])\hs{\rm
e}^{-\tau(\nu,r)},
\label{eq:pa1}
\end{equation} where $\tau(\nu,r)$ is the optical depth to photons {\it emitted} at frequency $\nu$ to a hydrogen atom at a distance $r$ from the central source. Furthermore, the factor $[1-Hr/c]$ accounts for the fact that the Hubble expansion redshifts the photons by a factor of $\Delta v=Hr$ over a distance $r$. Eq~\ref{eq:pa1} does not account for the fact that individual Ly$\beta$ photons are scattered multiple times, which modifies the actual value of mean intensity (see \S~\ref{sec:result}). 

For illustration purposes it is useful to rewrite Eq~\ref{eq:pa1} into
a form similar to Eq~\ref{eq:pscat}:
\begin{equation}
P_{\beta}(r)=\frac{L_{\nu}\hs f_{\beta}\hs \pi e^2}{4\pi r^2\hs
h\nu_{\beta}\hs m_e c}\times S(r),
\label{eq:pscat2}
\end{equation} where $S(r)$ (the 'suppression factor') is given by

\begin{equation}
S(r)=\frac{1}{\sqrt{\pi}}\int dx\hs \phi(x){\rm e}^{-\tau(x,r)}.
\label{eq:sr}
\end{equation} Here, $x$ denotes a dimensionless frequency variable of the form $x=(\nu-\nu_{\beta})/\Delta \nu_{D}$, where $\Delta \nu_{D}\equiv v_{\rm th}\nu_{\beta}/c$ and $v_{\rm th}=\sqrt{2kT/m_p}$. Furthermore, $\phi(x)$ is the Voigt function (see Eq~10.77 of Rybicki \& Lightman 1979. Note that this function obeys $\frac{1}{\sqrt{\pi}}\int dx\hs \phi(x)=1$. Hence, we recover Eq~\ref{eq:pscatb} when we set $\tau=0$.). The function $S(r)$ quantifies the inaccuracy in assuming that the absorbing gas is optically thin. 

\begin{figure}
\vbox{\centerline{\epsfig{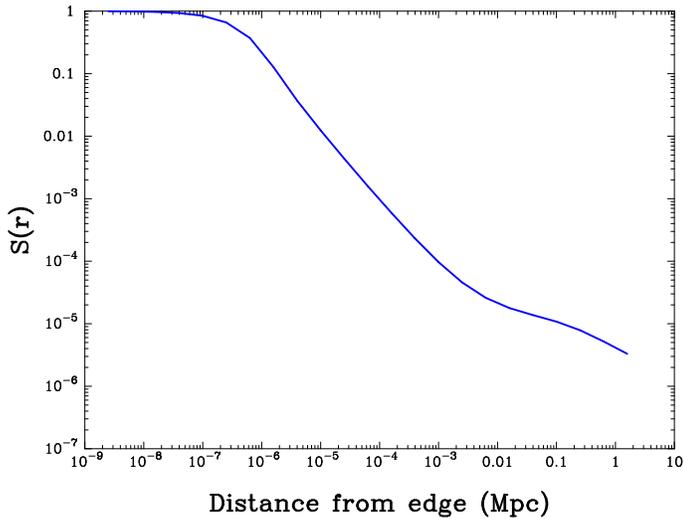}}}
\caption[]{This figure shows suppression factor $S(r)$, which
quantifies the reduction of the Ly$\beta$ scattering rate if one
assumes an e$^{-\tau}$ suppression of the Ly$\beta$ mean
intensity. Because of the large Ly$\beta$ absorption cross section the
suppression is strong:  $S(r)<10^{-2}$ for $r>10^{-5}$ Mpc (see text).}
\label{fig:lyarates}
\end{figure}

We again assume that the edge of the HII region is infinitely sharp
(as in \S~\ref{sec:detection}) and extends out to $r=R_{\rm HII}=5$
Mpc (see \S~\ref{sec:sim}, roman numeral III). The optical depth
$\tau(r,\nu)$ is then
\begin{equation}
\tau(\nu,r)=\int_{R_{\rm HII}}^r ds\hs n_{\rm
HI}(s)\sigma_{\beta}(\nu[1-Hs/c]).
\label{eq:tau}
\end{equation} We plot the suppression factor $S(r)$ in Figure~\ref{fig:lyarates}, which shows that $S(r)<10^{-2}$ for $r>10^{-5}$ Mpc. The reason that $S(r)$ drops so rapidly can be understood as follows: regardless of their frequency, photons traverse by definition an average optical depth $\tau=1$ into the neutral intergalactic medium before being absorbed. For Ly$\beta$ photons, this implies that the vast majority of photons are absorbed well before they have redshifted into resonance. That is, $\tau(x,r)\sim1$ when $x\gg 1$ and therefore $\phi(x) \ll 1$, in the notation of Eq~\ref{eq:sr}. Only those photons that have redshifted into resonance at the very edge can scatter near resonance (i.e. $\phi(x)=1$ while $\tau(x,r)\ll 1$, also see \S\ref{sec:result}). This special region is very thin: the line-center optical-depth of neutral intergalactic gas in the Ly$\beta$ line is $\tau_0\equiv \sigma_{\beta,0}n_{\rm HI}r=2.4\times 10^3(T_{\rm gas}/10^4)^{-1/2}(r/{\rm kpc})$. That is, the line center optical depth approaches unity at $r\sim 10^{-3}$ kpc=$10^{-6}$ Mpc. The scattering rate that is obtained by combining Eq~\ref{eq:pscat2}-\ref{eq:tau} plotted as the {\it dashed line} in Figure~\ref{fig:lybscatrate}.

\label{lastpage}
\end{document}